\documentclass[twocolumn]{aastex62}
\usepackage{verbatim}
\usepackage{gensymb}
\usepackage{graphicx}
\usepackage{array}
\usepackage{booktabs}
\usepackage{natbib}
\usepackage{bm}
\usepackage{amsmath}
\newcolumntype{+}{>{\global\let\currentrowstyle\relax}}
\newcolumntype{^}{>{\currentrowstyle}}

\newcommand{\NOTll}{\hskip 0.4mm \not \hskip -0.4mm \ll}

\begin{document}

\title{The Distribution and Excitation of CH$_3$CN in a Solar Nebula Analog}

\correspondingauthor{Ryan A. Loomis}
\email{rloomis@cfa.harvard.edu}

\author{Ryan A. Loomis}
\affil{Harvard-Smithsonian Center for Astrophysics, Cambridge, MA 02138}

\author{L. Ilsedore Cleeves}
\affil{Harvard-Smithsonian Center for Astrophysics, Cambridge, MA 02138}

\author{Karin I. {\"O}berg}
\affil{Harvard-Smithsonian Center for Astrophysics, Cambridge, MA 02138}

\author{Yuri Aikawa}
\affil{Department of Astronomy, University of Tokyo, 7-3-1 Hongo, Bunkyo-ku, Tokyo 113-0033, Japan}

\author{Jennifer Bergner}
\affil{Harvard-Smithsonian Center for Astrophysics, Cambridge, MA 02138}

\author{Kenji Furuya}
\affil{Center for Computational Sciences, University of Tsukuba, 1-1-1 Tennodai, Tsukuba, Ibaraki 305-8577, Japan}

\author{V.V. Guzman}
\affil{Joint ALMA Observatory, Alonso de Cordova 3107 Vitacura, Santiago de Chile, Chile}

\author{Catherine Walsh}
\affiliation{School of Physics and Astronomy, University of Leeds, Leeds LS2 9JT, UK}

\begin{abstract}
    Cometary studies suggest that the organic composition of the early Solar Nebula was rich in complex nitrile species such a CH$_3$CN. Recent ALMA detections in protoplanetary disks suggest that these species may be common during planet and comet formation, but connecting gas phase measurements to cometary abundances first requires constraints on formation chemistry and distributions of these species. We present here the detection of seven spatially resolved transitions of CH$_3$CN in the protoplanetary disk around the T-Tauri star TW Hya. Using a rotational diagram analysis we find a disk-averaged column density of N$_T$=1.45$^{+0.19}_{-0.15}\times10^{12}$~cm$^{-2}$ and a rotational temperature of T$_{rot}$=32.7$^{+3.9}_{-3.4}$~K. A radially resolved rotational diagram shows the rotational temperature to be constant across the disk, suggesting that the CH$_3$CN emission originates from a layer at z/r$\sim$0.3. Through comparison of the observations with predictions from a disk chemistry model, we find that grain-surface reactions likely dominate CH$_3$CN formation and that \textit{in situ} disk chemistry is sufficient to explain the observed CH$_3$CN column density profile without invoking inheritance from the protostellar phase. However, the same model fails to reproduce a Solar System cometary abundance of CH$_3$CN relative to H$_2$O in the midplane, suggesting that either vigorous vertical mixing or some degree of inheritance from interstellar ices occurred in the Solar Nebula.
    
\end{abstract}


\section{Introduction}a
    Observations of comets and meteorites show that the planet and comet forming midplane of the young Solar Nebula had a rich organic volatile composition \citep[e.g.,][]{Mumma_2011}. ALMA observations and the recent \textit{Rosetta} mission have both explicitly shown that comets are abundant in nitrile species such as HCN and CH$_3$CN \citep[e.g.,][]{Cordiner_2014, LeRoy_2015}. Tracing the chemistry of this family of organic molecules is of particular interest, as HCN and related nitriles are the starting point for the eventual synthesis of important bio-molecules such as glycine \citep[e.g.,][]{Bernstein_2002, Powner_2009, Powner_2010, Patel_2015}.

    In the protoplanetary disks where other planetary systems are just starting to form, the smallest nitriles CN and HCN have long been well-known \citep[e.g.,][]{Dutrey_1997, vanZadelhoff_2001}. Larger nitriles such as HC$_3$N have only been found more recently \citep{Chapillon_2012}, and ALMA is just now beginning to reveal more complex species such as CH$_3$CN \citep{Oberg_2015, Bergner_2018}. The provenance of these species in disks is unclear, however, and will play a role in setting their final abundances in the forming cometary bodies. Organics may be directly inherited from the chemically rich protostellar stage \citep[e.g.,][]{Jorgensen_2016}, formed \textit{in situ} in the disk \citep[e.g.,][]{Sakai_2014, Walsh_2014}, or both pathways may contribute. If inheritance from the protostellar stage dominates then nitrile abundances will likely be similar across disks within a given stellar association, while dominant \textit{in situ} formation may imply that cometary nitrile abundances will be highly disk dependent.
    
    Testing the origin of nitriles such as CH$_3$CN in disks will require constraining their abundances and formation routes, allowing comparison to predictions from the different inheritance scenarios. As CH$_3$CN can efficiently form through both gas-phase and grain-surface reactions, however, resolved observations of the disk abundance distribution are necessary to determine its dominant formation route. Thus far, observational constraints have been sparse \citep{Oberg_2015, Bergner_2018}. Based on the inferred abundance of CH$_3$CN in MWC 480, \cite{Oberg_2015} concluded that \textit{in situ} grain surface chemistry must play an important role. This same chemistry should produce CH$_3$CN the disk midplane, affecting the composition of forming comets and planetesimals. Better constraints on the distribution and excitation of CH$_3$CN in disks are therefore crucial to test this hypothesis and connect disk chemistry with cometary measurements.
        
    In this paper we present the detection of seven lines of CH$_3$CN in the disk around TW Hya. A well-studied, old \citep[$\sim$10 Myr; e.g.,][]{Kastner_1997, Weinberger_2013} T Tauri star, TW Hya hosts the closest \citep[59.5$\pm$1 pc;][]{Gaia_2016} protoplanetary disk, and is a good analog for the Solar Nebula \citep[0.8 M$_{\odot}$, spectral type K7; e.g.,][]{Rucinski_1983, Bergin_2013}. We present the observations and the details of their reduction and imaging in \S2. In \S3, we use a rotational diagram analysis to empirically constrain the CH$_3$CN column density and rotational temperature, both disk-averaged and radially resolved, and compare with predictions from detailed chemical models. In \S4 we discuss these results and their implications for midplane CH$_3$CN abundances and incorporation into planetesimals and forming comets. A summary is given in \S5.

\section{Observations}
    \subsection{Observational details}
        \begin{deluxetable*}{ccccccc}
        \tablecaption{Observed CH$_3$CN transitions}
        \tablecolumns{7}
        \tablewidth{\columnwidth}
        \tablehead{                                                                            
        	\colhead{Transition}          & \colhead{Symmetry}  & \colhead{Frequency}         & \colhead{E$_{u}$}   & \colhead{S$_{ij}\mu^{2}$}   & \colhead{Int. Flux Dens.$^{a}$}         & \colhead{Filter response} \\
        	                    &           & \colhead{(MHz)}             & \colhead{(K)}       & \colhead{(D$^2$)}           & \colhead{(mJy~km~s$^{-1}$)} & \colhead{($\sigma$)}}
        	\startdata
            12$_{0}$--11$_{0}$  & A         & 220747.3$^{b}$    & 68.9      & 183.7$^{c}$       & 82~$\pm$~7       & 17.8  \\
            12$_{1}$--11$_{1}$  & E         & 220743.0$^{b}$    & 76.0      & 182.5$^{c}$       & 78~$\pm$~7       & 15.9  \\
            12$_{2}$--11$_{2}$  & E         & 220730.3$^{b}$    & 97.4      & 178.6$^{c}$       & 41~$\pm$~7       & 7.2  \\
            \\
            13$_{0}$--12$_{0}$  & A         & 239137.9$^{b}$    & 80.3      & 199.1$^{c}$       & 81~$\pm$~7       & 13.1  \\
            13$_{1}$--12$_{1}$  & E         & 239133.3$^{b}$    & 87.5      & 197.9$^{c}$       & 70~$\pm$~7       & 9.3  \\
            13$_{2}$--12$_{2}$  & E         & 239119.5$^{b}$    & 108.9     & 194.3$^{c}$       & 28~$\pm$~7       & 5.7  \\
            13$_{3}$--12$_{3}$  & A         & 239096.5$^{b}$    & 144.6     & 188.5$^{c}$       & 12~$\pm$~7       & 3.5  \\
            \enddata
        \tablenotetext{}{$^{a}$ Velocity-integrated between 2.1--3.7~km~s$^{-1}$.}
        \tablenotetext{}{$^{b}$ Center frequency of collapsed hyperfine components (spacing smaller than channel width).}
        \tablenotetext{}{$^{c}$ S$_{ij}\mu^{2}$ of combined hyperfine components.}
        \end{deluxetable*}
        
        TW Hya was observed on 29-Dec-2016 and 09-Jan-2017 in Band 6 as part of the ALMA Cycle 4 project 2016.1.01046.S. The first execution block included 43 antennas with projected baseline lengths between 15 and 460~m (11--353~k$\lambda$). The second execution block included 47 antennas with projected baseline lengths between 15 and 384~m (11--295~k$\lambda$). The on-source integration times were 32 and 31 minutes, respectively, for a total on-source integration time of 63 minutes. The correlator setup was identical for both execution blocks and included a Time Division Mode (TDM) continuum window centered at 237~GHz with a bandwidth of 2~GHz as well as Frequency Division Mode (FDM) spectral windows centered at 219.560, 220.740, and 239.112~GHz. These spectral windows had bandwidths of 58.59~MHz and channel spacings of 61~kHz ($\sim$0.08~km s$^{-1}$), and they targeted the C$^{18}$O J=2--1, CH$_3$CN J=12--11, and CH$_3$CN J=13--12 molecular transitions, respectively. As CH$_3$CN is a prolate symmetric top with C$_{3v}$ symmetry its rotational spectrum has a k-ladder structure with two spin symmetry states (A/E), allowing a single set of observations to probe a wide range of upper state energies. We cover three transitions in the J=12--11 k-ladder and four transitions in the J=13--12 k-ladder, tabulated in Table 1. The 12$_{3}$--11$_{3}$ transition was not covered in our spectral setup.
        
        For both executions, the quasar J1058+1033 was used for bandpass calibration and the quasar J1037-2934 was used for phase calibration. Callisto was used as the flux calibrator for the first execution, and Ganymede was used as the flux calibrator for the second execution. We additionally used the disk continuum emission in each execution block to perform three rounds of phase self-calibration and one round of amplitude self-calibration in CASA version 4.3. These solutions were then applied to the spectral line observations.
    
    \subsection{Results}
        The observations were first analyzed using a matched filtering technique for identifying weak line emission, described in \cite{Loomis_2018}. The C$^{18}$O J=2--1 transition was imaged using \texttt{CLEAN} at the native spectral resolution (61~kHz, $\sim$0.08~km s$^{-1}$) with Briggs weighting (robust=0.5), producing a high SNR image cube (peak SNR=32). This image cube was then used as a filter for the CH$_3$CN spectral windows using the \texttt{VISIBLE} code\footnote{\texttt{VISIBLE is publicly available at \url{https://github.com/AstroChem/VISIBLE}}}. From the resultant filter impulse response spectra we detected all three transitions in the J=12--11 k-ladder covered by the spectral setup and all four transitions in the J=13--12 k-ladder with spectral coverage. The peak filter responses for each transition are given in Table 1.

        \begin{figure*}[ht!]
        \centering
        \includegraphics[width=0.65\textwidth]{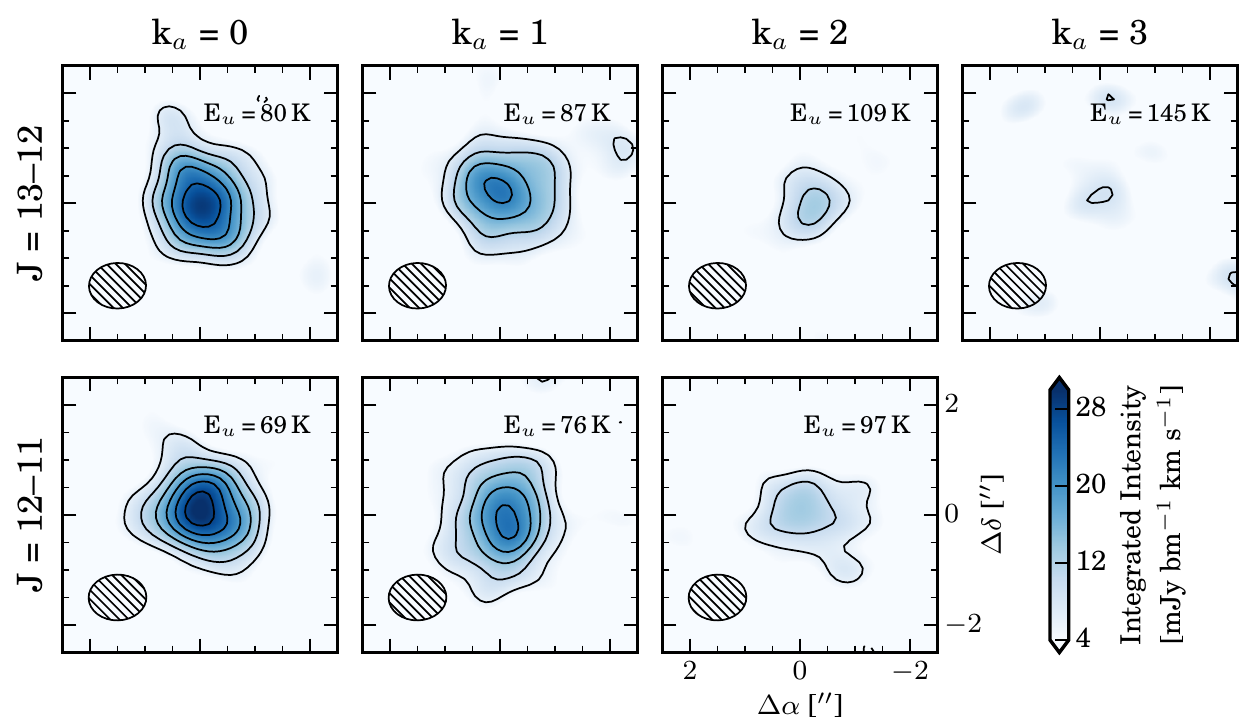}
        \caption{{\small Integrated intensity (moment-0) images of the observed CH$_3$CN transitions, velocity-integrated between 2.1--3.7~km~s$^{-1}$. All panels share the same intensity scale. Contours are [3,5,7,...]$\times\sigma$, where $\sigma$=2.2~mJy~beam$^{-1}$~km~s$^{-1}$. The synthesized beam is shown in the lower left of each panel.} 
        \label{Figure 1}}
        \end{figure*}

        The seven detected transitions were then individually imaged using \texttt{CLEAN} with natural weighting and a velocity resolution of 0.2~km~s$^{-1}$, centering each image cube on the transition rest frequency. The J=13--12 transitions had a small uv-taper (`outertaper' = 0$\farcs$35) applied to force the synthesized beam to match that of the J=12--11 transitions (1$\farcs$05 $\times$ 0$\farcs$83). The rms of the image cubes was $\sim$3.2~mJy~beam$^{-1}$ in each channel, and channel maps are presented in Appendix A. Moment-0 maps of the transitions are shown in Fig. \ref{Figure 1} and were created by integrating all emission between 2.1--3.7~km~s$^{-1}$ with no clipping threshold.

        \begin{figure}[ht!]
        \centering
        \includegraphics[width=0.9\columnwidth]{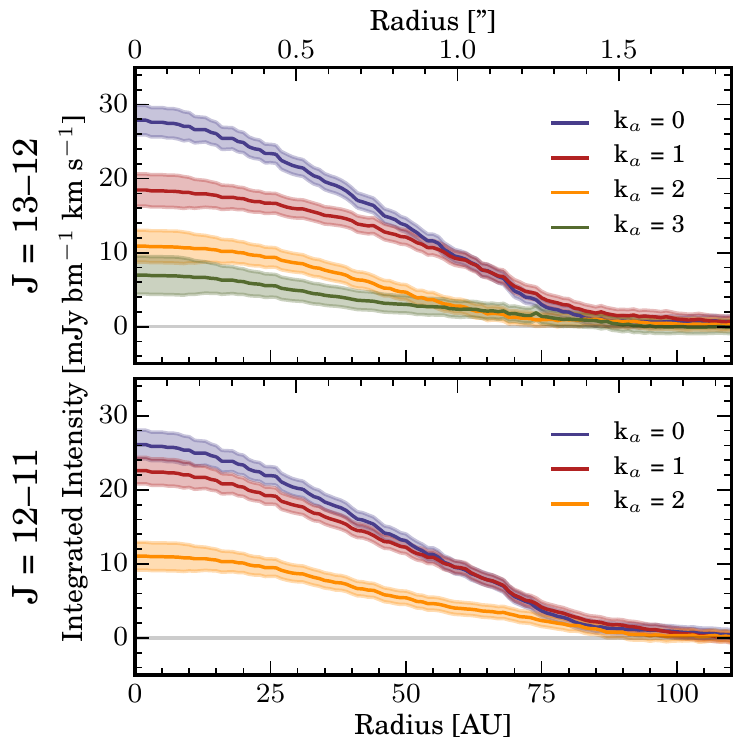}
        \caption{{\small Deprojected and azimuthally averaged radial intensity profiles of the observed CH$_3$CN transitions. Shaded regions denote 1$\sigma$ uncertainty levels, where the uncertainty in each radial bin was calculated by dividing the respective moment-0 image rms by the square root of the number of independent measurements in that bin (i.e. the bin circumference divided by the beam size).} 
        \label{Figure 2}}
        \end{figure}
        
        Deprojected and azimuthally averaged radial intensity profiles (Fig. \ref{Figure 2}) were calculated from the moment-0 maps in Fig. \ref{Figure 1} using an inclination of 7$\degree$ and PA of 155$\degree$ \citep{Qi_2004, Andrews_2012, Andrews_2016}. All transitions are centrally peaked, but the beam size is relatively large ($\sim$50-60~AU) compared to the extent of the emission, leaving open the possibility of a ringed morphology at small radii. The transitions all have similar profile shapes, with their relative strengths decreasing with increasing k$_a$.


        \begin{figure*}[ht!]
        \centering
        \includegraphics[width=0.65\textwidth]{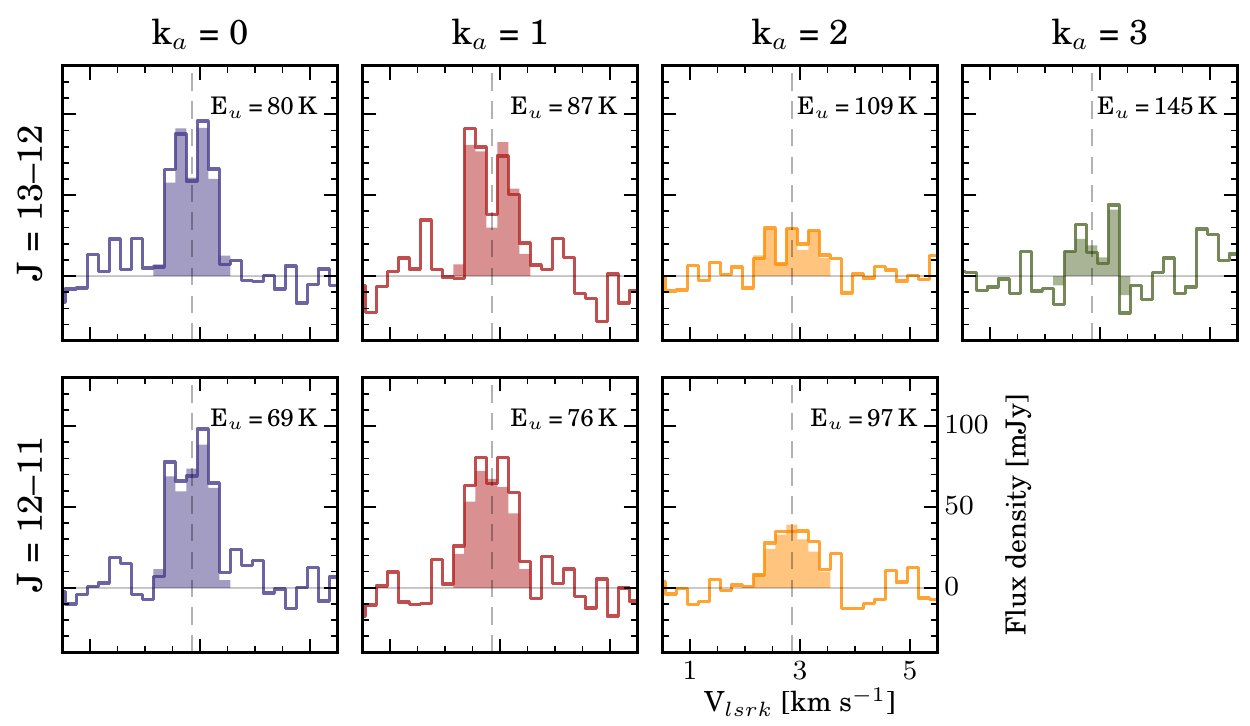}
        \caption{{\small Spectra of the observed CH$_3$CN transitions. Line profiles were extracted with an elliptical aperture mask 3\farcs5 in diameter and shaded profiles were extracted using a Keplerian mask convolved with the synthesized beam.} 
        \label{Figure 3}}
        \end{figure*}
        
        Spectra were extracted for each transition using two methods. First, an elliptical mask 3\farcs5 in diameter was used (line profiles in Fig. \ref{Figure 3}), which corresponds to a radial extent of $\sim$105~AU and encapsulates all emission given the radial profiles in Fig. \ref{Figure 2}. Second, a Keplerian mask convolved with the synthesized beam was used to extract the shaded profiles in Fig. \ref{Figure 3}, offering a better estimate of the true flux of each transition. The convolved Keplerian mask was truncated at a radial distance of 105~AU, and thus covers the same total solid angle as the elliptical mask. Flux measurements, listed in Table 1, were made by integrating the Keplerian extracted spectra between 2.1 and 3.7~km~s$^{-1}$. Uncertainty on each flux measurement was determined through bootstrapping, repeating the extraction and integration 10,000 times on an identical number of randomly-selected nearby emission-free channels (sampled with replacement). The standard deviation of these values is reported as the uncertainty on the flux measurement.

\section{CH$_3$CN column density and excitation temperature}
    The k-ladder structure of CH$_3$CN's rotational spectrum allows multiple transitions to be observed simultaneously, spanning a wide range of upper state energies. With this lever arm, the CH$_3$CN column density and excitation temperature can be well-constrained through a rotational diagram analysis \citep[e.g.,][]{Goldsmith_1999}. We initially assume local thermodynamic equilibrium (LTE) excitation, as the critical densities of the J=13--12 and J=12--11 CH$_3$CN transitions are $\sim$2.6$\times$10$^{6}$ and $\sim$2.0$\times$10$^{6}$~cm$^{-3}$, respectively, at a typical disk molecular layer temperature of 40K \citep[extrapolated to higher J and interpolated in temperature from][]{Shirley_2015}. Typical disk gas densities are $>$~1$\times$10$^{6}$~cm$^{-3}$, apart from the upper regions of the disk atmosphere (z/r$>$0.6), which we do not expect these observations to probe.

    \subsection{Disk-averaged analysis}
        We first calculate a disk-averaged column density and excitation temperature. Under an assumption of optically thin emission, the column density of molecules in the upper state of each transition, $N_{u}^{thin}$, is related to the emission surface brightness, $I_{\nu}$ through the equation:
        \begin{equation}
            I_{\nu} = \frac{A_{ul} N_{u}^{thin} h c}{4 \pi \Delta v}, 
        \end{equation}   
        where $A_{ul}$ is the Einstein coefficient and $\Delta v$ is the linewidth \citep[e.g.,][]{Bisschop_2008}. The disk-averaged emission intensity is $I_{\nu} = S_{\nu}/\Omega$, where $S_{\nu}$ is the flux density and $\Omega$ is the solid angle subtended by the source. Substituting for $I_{\nu}$ and inverting Eq. 1:
        \begin{equation}
            N_{u}^{thin} = \frac{4 \pi S_{\nu} \Delta v}{A_{ul} \Omega h c}.
        \end{equation}   
        $S_{\nu} \Delta v$ is the integrated flux density reported for each transition in Table 1, and we use the total solid angle covered by the beam-convolved Keplerian mask as an estimate of $\Omega$.
        
        Following \citet{Gordy_Cook}, the upper state level population $N_{u}$ can be related to the total column density $N_T$ by the Boltzmann equation:
        \begin{equation}
            \frac{N_{u}}{g_u} = \frac{N_T}{Q(T_{rot})}e^{-E_u / k T_{rot}},
        \end{equation}   
        where $g_u$ is the degeneracy of the upper state level, $Q$ is the molecular partition function, $T_{rot}$ is the rotational temperature, and $E_u$ is the upper state energy. CH$_3$CN is a symmetric top with C$_{3v}$ symmetry, and the upper state degeneracy $g_u$ can be written as
        \begin{equation}
            g_u = g_J g_K g_I
        \end{equation}   
        where $g_J = 2J+1$, $g_K = 1$ for $K = 0$, and $2$ for $K \neq 0$, and $g_I$ is the reduced nuclear spin degeneracy. For CH$_3$CN, $g_I$ can be defined as
        \begin{equation}
            g_I = 
            \begin{cases}
                \frac{1}{3} \left[1 + \frac{2}{(2I+1)^2} \right], & \text{for } K=0,3,6,\ldots \\
                \frac{1}{3} \left[1 - \frac{2}{(2I+1)^2} \right], & \text{for } K \text{ not divisible by 3}
            \end{cases}   
        \end{equation}
        The partition function $Q$ can be approximated for a molecule with C$_{3v}$ symmetry as
        \begin{equation}
            Q(T_{rot}) = \left(\frac{5.34\times10^6}{\sigma}\right)\left(\frac{T_{rot}^{3}}{B^2 A}\right)^{1/2},
        \end{equation}   
        where $\sigma$ is a unitless symmetry parameter, equal to 3 for a molecule with C$_{3v}$ symmetry, and A and B are the molecular rotational constants. Values for these rotational constants and all other spectral line data were taken from the Spectral Line Atlas of Interstellar Molecules (SLAIM)\footnote{Available at \url{http://www.splatalogue.net.}} \citep[F.J. Lovas, private communication,][]{Remijan_2007}. 
        
        In a conventional rotational diagram analysis \citep[e.g.,][]{Goldsmith_1999}, taking the logarithm of Eq. 3 allows for a linear least squares regression:
        \begin{equation}
            \ln \frac{N_{u}}{g_u} = \ln N_T - \ln Q(T_{rot}) - \frac{E_u}{k T_{rot}}.
        \end{equation}   
        If the level populations, $N_u/g_u$, are semi-log plotted against the upper state energies, $E_u$, then the rotational temperature, $T_{rot}$, and total column density, $N_T$, can be derived from the best fit slope and intercept, respectively. Under the assumption of optically thin emission, $N_{u}^{thin} = N_{u}$, Eq. 2 can be used to calculate $N_u/g_u$. The optical depth of the observed CH$_3$CN transitions is unknown \textit{a priori}, however. In the case that the optical depth $\tau \NOTll 1$, an optical depth correction factor $C_{\tau}$ must be applied:
        \begin{equation}
            C_{\tau} = \frac{\tau}{1-e^{-\tau}},
        \end{equation}   
        and thus the true level populations become
        \begin{equation}
            N_u = N_u^{thin} C_{\tau},
        \end{equation}   
        such that Eq. 7 is rewritten as
        \begin{equation}
            \ln \frac{N_{u}}{g_u} + \ln C_{\tau} = \ln N_T - \ln Q(T_{rot}) - \frac{E_u}{k T_{rot}}.
        \end{equation}   
        The optical depths of individual transitions are often directly determined through hyperfine ratios or observations of isotopomers, but can also be related back to the upper state level populations:
        \begin{equation}
            \tau_{ul} = \frac{A_{ul}c^3}{8 \pi \nu^3 \Delta v} N_u (e^{h\nu / k T_{rot}} - 1).
        \end{equation}   
        $C_{\tau}$ can therefore be written as a function of $N_u$ and substituted into Eq. 10 to construct a likelihood function $L(N_u, T_{rot})$ which can then be used for $\chi^2$ minimization.
        
        \begin{figure}[ht!]
        \centering
        \includegraphics[width=\columnwidth]{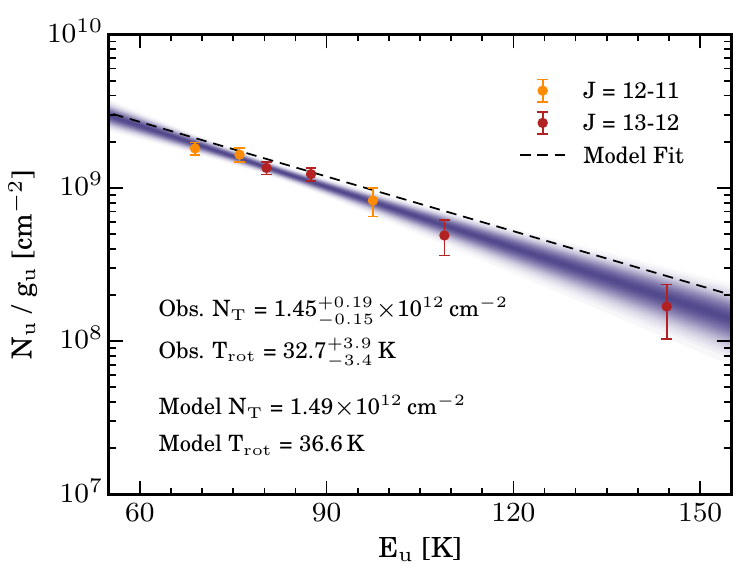}
        \caption{{\small CH$_3$CN rotational diagram, constructed using disk averaged intensities. J=12--11 and J=13--12 transitions are shown in orange and red, respectively. Random draws from the fit posteriors are plotted in blue and a fit to simulated observations from the chemical model described in \S3.3 is plotted in dashed black.} 
        \label{Figure 4}}
        \end{figure}
        
        Given this likelihood function, we use the affine-invariant Markov chain Monte Carlo (MCMC) code \texttt{emcee} \citep{Foreman-Mackey_2013} to fit the data and generate posterior probability distributions of both $N_u$ and $T_{rot}$ (see Fig. \ref{corner fig} in Appendix B). These probability density functions describe the range of possible column densities and rotational temperatures that are consistent with our observed data. Random draws from these posteriors are plotted in blue in Fig. \ref{Figure 4}, with $\tau$ corrected values of $N_u/g_u$ plotted against $E_u$. We find a disk-averaged column density of N$_T$=1.45$^{+0.19}_{-0.15}\times10^{12}$~cm$^{-2}$ and a rotational temperature of T$_{rot}$=32.7$^{+3.9}_{-3.4}$~K, where parameters and uncertainties are listed as the 50th, 16th, and 84th percentiles from the marginalized posterior distributions, respectively. Corresponding values of $\tau$ range between 0.002--0.012, confirming that these transitions of CH$_3$CN are optically thin. These values show a good fit to both the complete dataset as well as the individual J=12--11 and J=13--12 k-ladders (shown in orange and red, respectively), consistent with the assumption of LTE excitation.
        
    \subsection{Radially resolved analysis}
\begin{figure}[ht!]
        \centering
        \includegraphics[width=\columnwidth]{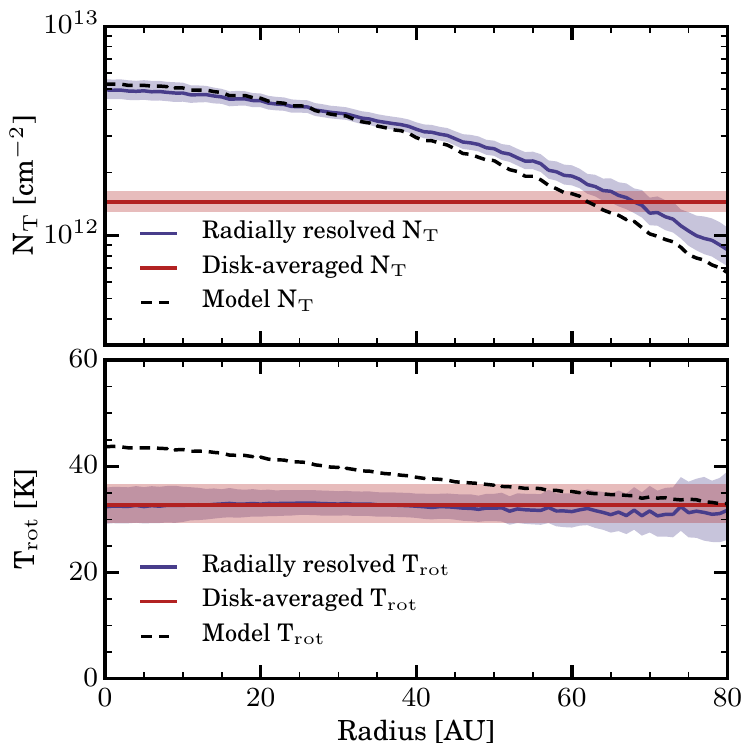}
        \caption{{\small Radial profiles of the fit CH$_3$CN column density (upper panel) and rotational temperature (lower panel). Best fit values and 1$\sigma$ uncertainties are plotted in blue for the radially resolved analysis and in red for the disk-averaged analysis. Fits to the simulated observations described in \S3.3 are plotted in dashed black.} 
        \label{Figure 5}}
        \end{figure}

        As the observed CH$_3$CN transitions are strongly detected and moderately resolved (with a beam size of $\sim$50~AU),  N$_T$ and T$_{rot}$ can be further constrained as a function of radius. We repeat the rotational diagram analysis previously described, but now use intensities from the radial profiles of each transition from Fig. \ref{Figure 2} rather than disk-averaged intensities. Posterior distributions for N$_T$ and T$_{rot}$ are calculated at intervals of 1~AU and are plotted in Fig. \ref{Figure 5}. The rotational diagrams (not shown) remain log-linear and well-behaved out to $\sim$70~AU but become non-linear exterior to this distance, leading to the large uncertainties in T$_{rot}$.
        
        The observed N$_T$ profile decreases with radius from 5 to 0.9$\times10^{12}$~cm$^{-2}$. This is consistent with the disk-averaged column density of N$_T$=1.45$^{+0.19}_{-0.15}\times10^{12}$~cm$^{-2}$, which is overplotted in red in Fig. \ref{Figure 5}. The disk-averaged column density is biased towards the low end of the radially resolved column density range. The majority of the emission (and therefore molecular column) is concentrated in the inner regions of the disk (R$\lesssim$50~AU) and the disk-averaged intensities (integrated out to R=105~AU) are therefore diluted. T$_{rot}$ shows a flat radial profile, ranging between 30--34~K, consistent with the disk-averaged rotation temperature, T$_{rot}$=32.7$^{+3.9}_{-3.4}$~K.
    
    \subsection{Comparison to chemical models}
\begin{figure}[ht!]
        \centering
        \includegraphics[width=\columnwidth]{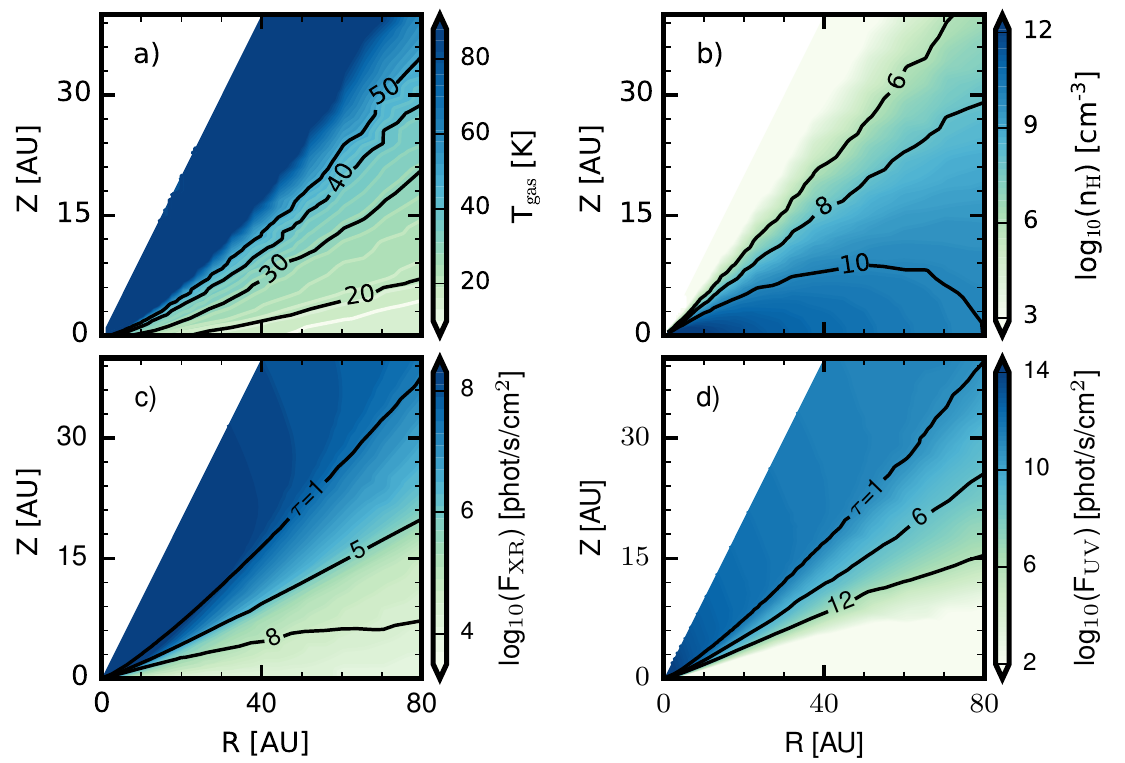}
        \caption{{\small Assumed physical structure of TW Hya used for chemical modeling, taken from \citep{Cleeves_2015}. \textit{Panels (a,b):} Disk temperature and density structures, respectively. \textit{Panels (c,d):} X-ray and FUV radiation fields, respectively, with optical depths overplotted as contours.} 
        \label{Figure 6}}
        \end{figure}

        We compare the empirical contraints derived in \S3.1 and \S3.2 to the predictions of a time-dependent chemical model \citep{Fogel_2011, Cleeves_2014} evolved for 1~Myr. The assumed density and temperature structures, constrained by the TW Hya SED and previous HD observations \citep{Bergin_2013} in \cite{Cleeves_2015}, are shown in Fig. \ref{Figure 6} panels a and b, respectively. The initial chemical abundances of the model, listed in Table 2, are based on values from \cite{Cleeves_2015} but with updated CO and H$_2$O depletion factors. CO is depleted by a factor of 20 to approximately compensate for the known carbon depletion in TW Hya \citep[e.g.][]{Favre_2013, Kama_2016, Schwarz_2016}. H$_2$O is depleted in the model by a factor of 100 \citep[e.g.][]{Du_2015}. No CH$_3$CN is included in the initial abundances, and thus all CH$_3$CN in the model is produced \textit{in situ}. The FUV and X-ray radiation fields within the disk (Fig. \ref{Figure 6}, panels c,d) were calculated using the Monte Carlo code and cross sections from \cite{Bethell_2011}, the observed TW Hya FUV spectrum \cite{Herczeg_2002, Herczeg_2004}, and a best-fit X-ray model for TW Hya from \cite{Cleeves_2015}. A reduced cosmic ray ionization rate was assumed, as TW Hya has been found to have a reduced cosmic ray ionization rate due to exclusion either by winds or magnetic fields \citep[$\zeta_{\rm CR} \sim 2\times10^{-19}$ s$^{-1}$; SSX model;][]{Cleeves_2015}.

        \begin{deluxetable}{cccc}
        \tablecaption{Chemical model initial abundances}
        \tablecolumns{4}
        \tablewidth{\columnwidth}
        \tablehead{
        	\colhead{Species} & \colhead{Abundance$^a$} &  \colhead{Species}  & \colhead{Abundance$^a$}}
        	\startdata
        	H$_2$       & $5.00\times10^{-1}$   & He            & $1.40\times10^{-1}$ \\
        	N$_2$       & $3.75\times10^{-5}$   & CO            & $7.00\times10^{-6}$ \\
        	H$_2$O(gr)  & $2.50\times10^{-6}$   & H$_{3}^{+}$   & $1.00\times10^{-8}$ \\
        	HCO$^+$     & $9.00\times10^{-9}$   & C$_2$H        & $8.00\times10^{-9}$ \\
        	CS          & $5.00\times10^{-9}$ 	& SO            & $4.00\times10^{-9}$ \\
        	C$^+$       & $1.00\times10^{-9}$   & Si$^+$        & $1.00\times10^{-11}$ \\
        	Mg$^+$      & $1.00\times10^{-11}$ 	& Fe$^+$        & $1.00\times10^{-11}$ \\
            \enddata
        \tablenotetext{}{$^{a}$ Abundances are relative to the proton density $n_p = 2 n_{H_{2}}$.}
        \end{deluxetable}
        
        The chemical reaction network contains a total of 5970 reactions and 600 species. Within this network, three reactions are primarily responsible for the formation of CH$_3$CN \citep{Walsh_2014, Wakelam_2015}. In the gas phase, formation occurs through the radiative association reaction \citep{Herbst_1985}
        \begin{equation}
            \mathrm{CH_3^+} + \mathrm{HCN} \rightarrow \mathrm{CH_3CNH^+} + h\nu,
        \end{equation}   
        followed by dissociative recombination
        \begin{equation}
            \mathrm{CH_3CNH^+} + \mathrm{e^-} \rightarrow \mathrm{CH_3CN} + \mathrm{H}.
        \end{equation}   
        It should be noted, however, that the implicit isomerization of CH$_3$NCH$^+$ to CH$_3$CNH$^+$ in Eq. 12 as written would likely require a three-body interaction to be efficient \citep[e.g.][]{Anicich_1994} and is not firmly established in the literature as a viable process at the low densities present in protoplanetary disks. An analogous reaction for HNC is also possible
        \begin{equation}
            \mathrm{CH_3^+} + \mathrm{HNC} \rightarrow \mathrm{CH_3CNH^+} + h\nu,
        \end{equation}   
        with a disk-integrated HNC/HCN ratio of $\sim$0.1-0.2 having been previously measured for TW Hya \citep{Graninger_2015}. Destruction pathways for gas-phase CH$_3$CN include UV photodissociation into CH$_3$ + CN and reactions with C$^+$.
        
        On grain surfaces, there are two viable formation pathways through a Langmuir-Hinshelwood mechanism: (1) sequential hydrogenation of C$_2$N or (2) a neutral-neutral grain surface reaction between CH$_3$ and CN \citep{Wakelam_2006, Walsh_2014}. A reactive desorption efficiency of 1\% and a photodesorption yield of 10$^{-3}$ were assumed, with an additional assumption that the CH$_3$CN molecule always desorbs intact. The validity of these assumptions and their impact are discussed in more detail in \S4.1.3. Thermal desorption and freeze-out in the model are treated using the Polyani-Wigner relation, with an assumed binding energy of 4680~K for CH$_3$CN \citep{Collings_2004}.
        
\begin{figure}[ht!]
        \centering
        \includegraphics[width=\columnwidth]{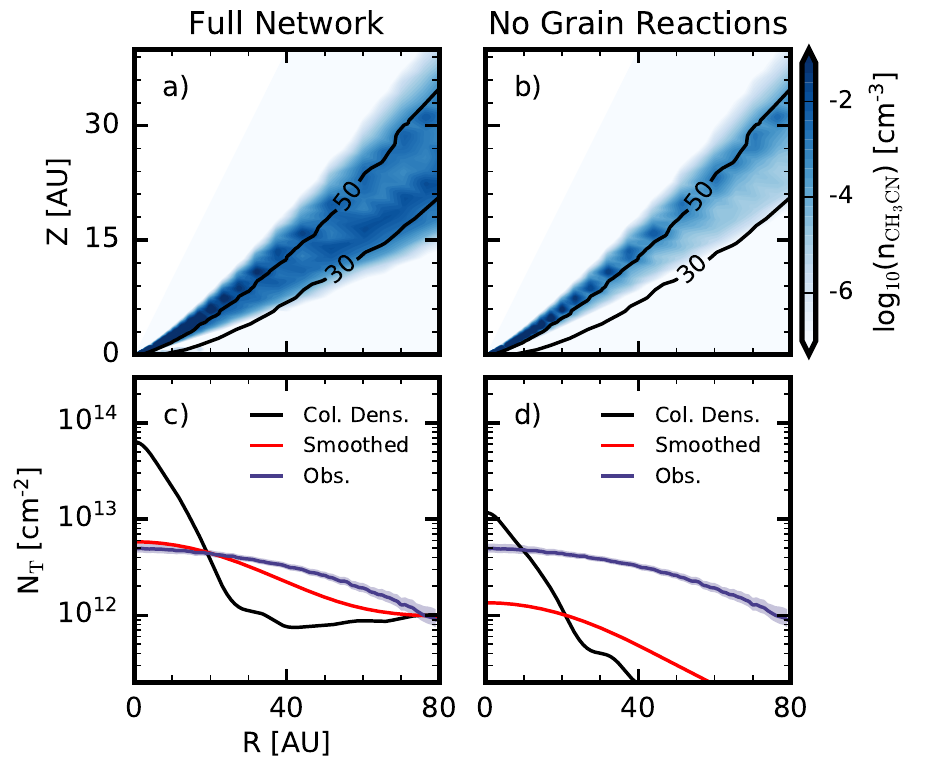}
        \caption{{\small \textit{Panels (a,b):} CH$_3$CN gas-phase abundances with grain-surface reactions turned on and off in the model, respectively. Temperature contours of 30 and 50~K are overlaid in black.} \textit{Panels (c,d):} CH$_3$CN gas-phase column densities for the abundance profiles shown in panels a and b, respectively. Column density profiles smoothed with the synthesized beam are overplotted in red, and the observed column density profile from Fig. \ref{Figure 5} is overplotted in blue.
        \label{Figure 7}}
        \end{figure}

        To isolate the respective contributions of gas phase and grain-surface formation mechanisms, we ran the chemical model twice, once with grain-surface reactions turned on and once with them turned off. Fig. \ref{Figure 7} shows the resultant gas phase CH$_3$CN abundance profiles (panels a,b). Both gas phase and grain-surface reactions contribute to the total CH$_3$CN reservoir, but form distinct vertical layers. Gas phase reactions produce CH$_3$CN in a layer at z/r $\sim$ 0.5, where the gas temperature is $\sim$50~K. The upper boundary of this layer sits along the FUV $\tau$=1 surface, and is primarily set by the UV photo-dissociation of CH$_3$CN. The lower boundary is set where the reactant CH$_3^+$ is no longer formed in appreciable quantities due to the FUV and X-ray optical depths. Grain-surface reactions meanwhile produce a layer of CH$_3$CN in the gas phase at z/r $\sim$ 0.3 (T$_{gas}$ $\sim$ 35~K), with formation dominated by sequential C$_2$N hydrogenation. C$_2$N is primarily formed in the gas phase through the reaction
        \begin{equation}
            \mathrm{C_{2}H} + \mathrm{N} \rightarrow \mathrm{H} + \mathrm{C_{2}N},
        \end{equation}   
        and then freezes out onto grain surfaces. The upper boundary of the grain-surface CH$_3$CN layer is set by this freeze-out of C$_2$N and the lower boundary is set by high optical depths limiting photodesorption of CH$_3$CN off the grain surfaces.
        
        The model abundance profiles were vertically integrated to calculate column density profiles (Fig. \ref{Figure 7} panels c,d). The model column density profiles, shown in black, were then convolved with the synthesize beam of the observations to produce the smoothed profiles shown in red. From these profiles, it is clear that grain-surface contributions in the full network model increase the integrated column density by a factor of 4--10 across the disk. Comparing the beam smoothed model profiles with the observed column density profile in Fig. \ref{Figure 5}, we find that gas phase reactions alone are insufficient to reproduce the observed column densities, while inclusion of grain-surface reactions reproduces the observed column densities within a factor of 2 across the disk.
        
        Although the beam smoothed profiles allow a rough comparison to our best fit observed column density profile, a more detailed comparison requires an identical analysis procedure for both the observations and chemical model results. We therefore used the chemical model output to calculate simulated emission profiles for the observed CH$_3$CN transitions using the radiative transfer code RADMC-3D \citep{Dullemond_2013}. A distance of 59.5~pc, PA of 155$\degree$, and stellar mass of 0.8 M$_{\odot}$ (to determine line broadening) were assumed for the radiative transfer \citep{Kastner_1997, Qi_2004, Andrews_2012, Andrews_2016}. An inclination of 8$\degree$ was assumed, which approximately accounts for the slight warp in the TW Hya disk \citep{Rosenfeld_2012} and was found to fit our observations relatively well (see Appendix A). Simulated ALMA observations were then calculated for each transition using the \texttt{vis$\textunderscore$sample} package \citep{Loomis_2018} and the antenna configuration of the original observations. 
        
        From these simulated observations, we repeated the analysis described in \S3.1 and calculated a disk-averaged rotational temperature and column density, overplotted in dashed black in Fig. \ref{Figure 4}. The disk-averaged calculated column density of N$_T$=1.49$\times$10$^{12}$~cm$^{-2}$ and rotational temperature of T$_{rot}$=36.6~K both agree with the observed column density of N$_T$=1.45$^{+0.19}_{-0.15}\times$10$^{12}$~cm$^{-2}$ and rotational temperature of T$_{rot}$=32.7$^{+3.9}_{-3.4}$~K within the errors.
        
        Similarly, we extracted deprojected and azimuthally-averaged radial intensity profiles and repeated the analysis described in \S3.2 to calculate resolved column density and rotational temperature profiles, overplotted in dashed black in Fig. \ref{Figure 5}. Both profiles match the observations relatively well. The model rotational temperature profile is up to $\sim$15~K warmer than the observations however, especially at radii $\lesssim$50~AU. This might be expected given the distribution of CH$_3$CN seen in Fig. \ref{Figure 7}, panel a, peaking at small radii and in a temperature layer $>$~50~K. This point is discussed further in \S4.1.

\section{Discussion}
    \subsection{CH$_3$CN abundance structure and formation chemistry}
        \subsubsection{Insights from observations}
            A rotational diagram analysis of our observations shows that CH$_3$CN in TW Hya emits at a near constant temperature of $\sim$30--35~K across the disk. From our assumed physical model of TW Hya, this temperature suggests emission from a vertical layer at z/r$\sim$0.3. These first observational constraints on the vertical distribution of CH$_3$CN are in good qualitative agreement with the layered CH$_3$CN distribution predicted by the chemical models in \cite{Oberg_2015}. Similarly, we find that our observed radial column density profile is in good qualitative agreement with the predictions of \cite{Walsh_2014} and the observational results of \cite{Oberg_2015}, which both found column density profiles between $\sim$10$^{12}$--10$^{13}$~cm$^{-2}$ that monotonically decreased with radius.
        
        \subsubsection{Comparison of chemical models and observations}
            We attempted to gain an intuition for the dominant CH$_3$CN formation pathway by comparing these observational results with two chemical models, with and without grain-surface chemistry. The full chemical network predicts emission which is in remarkably good agreement with our observations, especially given that the model has not been adjusted in any manner to match the data. The model with no grain-surface reactions underpredicts our measured fluxes by over an order of magnitude, suggesting that grain-surface formation of CH$_3$CN may be the main \textit{in situ} formation pathway. We additionally find that at all times in the full-network model, the total grain-surface formation rate dominates over the gas-phase formation rate by factors of $\sim$2-10. Dominant grain-surface formation is further supported by our observed temperature layer ($\sim$30--35~K, z/r$\sim$0.3) being better matched to the grain-surface formation layer than the gas-phase formation layer in Fig. \ref{Figure 7}. The limited spatial resolution of our observations and the caveats of our chemical model presented in \S4.1.3, however, prevent a more robust quantitative analysis of the relative gas-phase and grain-surface contributions.
        
            Although the full network model predicts adequate integrated fluxes, some differences remain between the model and observations. First, the radial profile of the model column density is slightly more centrally peaked (even after beam convolution) than the observed column density profile (Figs. \ref{Figure 5} and \ref{Figure 7}). Second, although the best-fit disk-averaged column density and rotational temperature are well-matched between the model and observations, the radially resolved model temperature profile is up to 15~K warmer than the observed temperature profile. 
            
            These differences are inherently linked; a more centrally peaked column density results in more emission at small radii, where the gas temperature is higher for a given z/r. A number of phenomena therefore could possibly explain these discrepancies. First, a central depletion in the CH$_3$CN emission cannot be ruled out by our observations, given their relatively low spatial resolution. By stacking the observations and examining the resultant channel maps, we are able to constrain the possible radial extent of such a feature to be less than 16~AU (see Appendix A). Second, the chemical model may over-estimate gas-phase CH$_3$CN production, resulting in an enhanced contribution of warm ($\sim$50~K) CH$_3$CN at small radii. Third, the temperature of the CH$_3$CN emission is likely sensitive to our assumptions about the physical structure of TW Hya and its FUV and X-ray radiation fields in particular, as the boundaries of the CH$_3$CN layer are directly linked to the optical depths of the radiation fields (see \S3.3). High resolution observations of smaller molecules such as HCN may allow future model refinement by anchoring the assumed initial conditions and disk physical characteristics.
        
        \subsubsection{Chemical model assumptions and caveats}
            A number of assumptions made in our chemical modeling complicate our interpretation of both the observations and models. First, molecules larger than CH$_3$CN are not included in the model and CH$_3$CN likely acts as a chemical `sink', enhancing model abundances. Second, given the uncertainties associated with the dominant gas-phase reaction (see \S3.3), it is unclear to what extent this pathway contributes to the observed CH$_3$CN abundance. If the assumed efficiency of this reaction in the model is too high, this may partially explain the higher rotational temperature found for the model compared to the data. Third, we assumed that CH$_3$CN is always able to photodesorb intact from grain-surfaces. Recent investigation on CH$_3$OH photodesorption suggests that larger molecules such as CH$_3$CN fragment and thus may have difficulty efficiently photodesorbing intact from grain surfaces \citep{Bertin_2016, Cruz-Diaz_2016, Walsh_2017}. If this is the case, reactive desorption may play a larger role as a mechanism for non-thermal CH$_3$CN desorption. Finally, our model initial conditions assume flat depletion of CO and H$_2$O across the disk. In reality, spatial variations in depletion and sequestration will result in a modified C/O ratio, which in turn will affect CH$_3$CN abundances. In particular, the formation of cyanides such as CH$_3$CN has been shown to be sensitive to carbon and oxygen abundances, with an enhanced C/O ratio resulting in more efficient cyanide formation \citep{Du_2015}. The expected nitrile enhancements for older disks with grain growth and radial drift have been tentatively observed by \cite{Guzman_2017}, and a similar effect was invoked by \cite{Bergin_2016} to explain hydrocarbon rings around TW Hya.
    
    \subsection{Implications for cometary CH$_3$CN abundances}
\begin{figure}[ht!]
        \centering
        \includegraphics[width=\columnwidth]{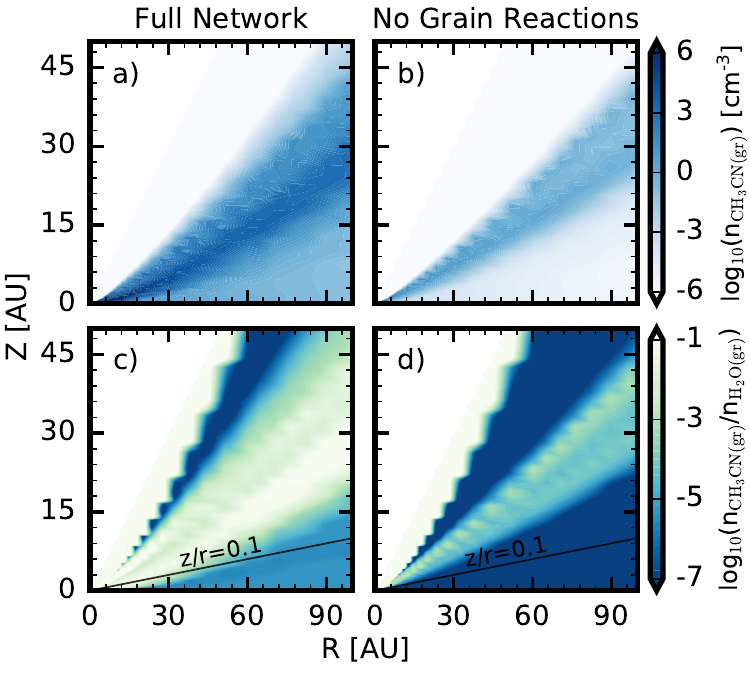}
        \caption{{\small \textit{Panels (a,b):} CH$_3$CN abundances in the solid-phase on grain-surfaces, calculated in chemical models with grain-surface reactions turned on and off, respectively. \textit{Panels (c,d):} CH$_3$CN(gr)/H$_2$O(gr) abundance ratios, calculated from the abundance profiles shown in panels a and b. Note that all CH$_3$CN in the models is formed \textit{in situ}, while initial abundances of H$_2$O(gr) are inherited from the protostellar phase.} 
        \label{Figure 8}}
        \end{figure}

        Our observations probe gas-phase abundances at a vertical layer in the disk of z/r$\sim$0.3. In contrast, comets form in the disk midplane and their bulk compositions are primarily set by grain-surface chemical abundances, rather than gas-phase abundances. Interpreting the implications of our observations for the chemical composition of comets therefore requires extrapolation to the disk midplane through our chemical model. Fig. \ref{Figure 8} panels a,b show the grain-surface abundances of CH$_3$CN in our chemical models with grain-surface chemistry turned on and off, respectively. Although no CH$_3$CN is formed on grain-surfaces in the latter model, freeze-out still results in a non-negligible CH$_3$CN grain-surface abundance. 
        
        To compare these grain-surface abundances to measured cometary CH$_3$CN abundances in the Solar System \citep[$\sim$10$^{-4}$ relative to H$_2$O, e.g.,][]{Mumma_2011}, panels c,d of Fig. \ref{Figure 8} show CH$_3$CN(gr)/H$_2$O(gr) abundance ratios across the disk model. Initial H$_2$O(gr) abundances in the midplane are inherited from the protostellar phase. We correct for the depletion factor assumed for gas-phase H$_2$O in the disk surface \citep[e.g.]{Du_2015}(see \S3.3), as ices in the disk midplane are not expected to be depleted. Gas-phase reactions alone (Fig. \ref{Figure 8}, panel d) are clearly insufficient to reproduce cometary CH$_3$CN abundances 7$\times$10$^{-5}$--3$\times$10$^{-4}$ \citep[e.g.,][]{Mumma_2011} near the midplane (ie z/r$<$0.1). Incorporation of grain-surface reactions (Fig. \ref{Figure 8}, panel c), however, produces abundances closer to cometary values (up to 5$\times$10$^{-4}$) in the comet forming regions of the disk (R$\lesssim$10--30~AU), consistent with the results of \cite{Walsh_2014}.

\begin{figure}[ht!]
        \centering
        \includegraphics[width=\columnwidth]{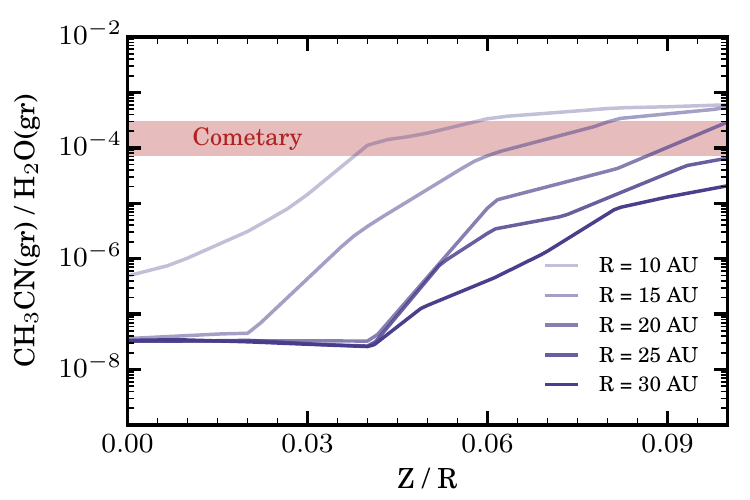}
        \caption{{\small CH$_3$CN(gr)/H$_2$O(gr) abundance ratios from Fig. \ref{Figure 8}, panel c (full chemical network) as a function of z/r, taken at different radial slices. The range of values for Solar System cometary abundances is shown in shaded red.} 
        \label{Figure 9}}
        \end{figure}

        Fig. \ref{Figure 9} plots CH$_3$CN(gr)/H$_2$O(gr) ratios as a function of z/r at a variety of radii in the disk, comparing these values to the range of known cometary CH$_3$CN(gr)/H$_2$O(gr) ratios. \textit{In situ} formation in the comet forming zone (R$\lesssim$10--30~AU) is insufficient to produce cometary abundances of CH$_3$CN at the midplane, but can easily yield these abundances at slightly higher disk layers (z/r$>$0.04), especially at smaller radii. A detailed understanding of the coupling between chemistry and vertical motion of material within the disk will be necessary to determine if CH$_3$CN produced higher in the disk can be efficiently transported to the midplane for incorporation into comets \citep[e.g.,][]{Semenov_2011, Furuya_2014}. In particular, such an analysis would require chemical modeling which incorporates both dust settling and turbulent diffusion, as these phenomena have pronounced effects on CH$_3$CN abundance distributions and the coupling between gas-phase and grain surface abundances \citep[e.g.,][]{Semenov_2011, Oberg_2015}.
        
        We note that the CH$_3$CN(gr)/H$_2$O(gr) ratios shown in Fig. \ref{Figure 9} are lower limits, as no inheritance from the protostellar stage is included in our models. Chemical modeling in \cite{Eistrup_2016} suggests that such inheritance is possible, with interstellar ices abundances preserved in environments which are shielded from cosmic rays \citep[e.g., TW Hya,][]{Cleeves_2015}. Observational evidence for protostellar inheritance of CH$_3$CN is mixed, however. \cite{Oberg_2015} find CH$_3$CN/HC$_3$N/HCN ratios in MWC 480 that are inconsistent with those found in protostellar regions such as IRAS 16293-2422 \citep[e.g.,][]{vanDishoeck_1995}, while \cite{Bergner_2018} find CH$_3$CN/HC$_3$N ratios in a number of disks which are consistent with protostellar values. 

        The detection of CH$_3$CN around TW Hya offers an additional opportunity to evaluate the possibility of protostellar inheritance, as this is the only disk where CH$_3$OH has been detected thus far \citep{Walsh_2016}. We find an approximate CH$_3$CN/CH$_3$OH column density ratio of unity, which is substantially higher than the few percent found in comets and around protostars \citep{Mumma_2011, Bergner_2017}. As discussed in \cite{Bergner_2018}, two scenarios could explain this finding: a higher photodesorption efficiency for CH$_3$CN than CH$_3$OH (where both species could either be inherited from the protostellar stage or form through \textit{in situ} grain surface chemistry), or gas phase production of nitriles such as CH$_3$CN could be enhanced by a high C/O ratio as discussed in \S4.1.3. Thus although our observations are inconsistent with preserved interstellar abundance ratios, it is possible that inheritance contributes to the total CH$_3$CN abundance in TW Hya.

\section{Summary}
    In summary, we have detected emission from seven transitions of CH$_3$CN toward TW Hya. A disk-averaged rotational analysis finds a column density of N$_T$=1.82$^{+0.25}_{-0.19}\times10^{12}$~cm$^{-2}$ and a rotational temperature of T$_{rot}$=29.3$^{+3.2}_{-2.8}$~K, and a radially resolved analysis shows this temperature to be flat across the disk. We interpret these results to suggest that CH$_3$CN emission originates from a layer at z/r$\sim$0.3 throughout the disk. Comparing these observations with the results of a disk chemistry model, we suggest that grain-surface reactions likely dominate CH$_3$CN formation. \textit{In situ} formation in the model is sufficient to explain observed CH$_3$CN fluxes, although further model refinement is necessary to accurately reproduce CH$_3$CN radial and vertical abundance profiles. Finally, we examine the CH$_3$CN(gr)/H$_2$O(gr) ratio predicted by our model and find that cometary abundances of CH$_3$CN are not present in the disk midplane, but can be found in slightly higher disk layers (z/r$>$0.04), suggesting that inheritance, dust settling, turbulent mixing, or a combination of these effects is necessary to replicate cometary CH$_3$CN abundances in the disk midplane.

\acknowledgments
We thank Edwin Bergin, Jamila Pegues, and Richard Teague for helpful conversations about the data analysis and CH$_3$CN chemistry. R.A.L. gratefully acknowledges funding from NRAO Student Observing Support. L.I.C. acknowledges the support of NASA through Hubble Fellowship grant HST-HF2-51356.001-A awarded by the Space Telescope Science Institute, which is operated by the Association of Universities for Research in Astronomy, Inc., for NASA, under contract NAS 5-26555. V.V.G. acknowledges support from the National Aeronautics and Space Administration under grant No. 15XRP15$\_$20140 issued through the Exoplanets Research Progam. K.I.\"O. acknowledges funding from the Simons Collaboration on the Origins of Life (SCOL). C.W. acknowledges financial support from the University of Leeds. The National Radio Astronomy Observatory is a facility of the National Science Foundation operated under cooperative agreement by Associated Universities, Inc.  This paper makes use of the following ALMA data: ADS/JAO.ALMA $\#$2016.1.01046.S. ALMA is a partnership of ESO (representing its member states), NSF (USA) and NINS (Japan), together with NRC (Canada) and NSC and ASIAA (Taiwan), in cooperation with the Republic of Chile. The Joint ALMA Observatory is operated by ESO, AUI/NRAO and NAOJ.

\software{Astropy \citep{Astropy_2013}, CASA \citep{McMullin_2007}, casa-python, Matplotlib \citep{Hunter_2007}, NumPy \citep{Jones_2001}, RADMC-3D \citep{Dullemond_2013}, SciPy \citep{vanderWalt_2011}, vis$\textunderscore$sample \citep{Loomis_2018}, VISIBLE \citep{VISIBLE_Zenodo, VISIBLE_ASCL}}

\clearpage

\appendix
\section{Channel Maps}
\begin{figure*}[ht!]
    \centering
    \includegraphics[width=0.8\textwidth]{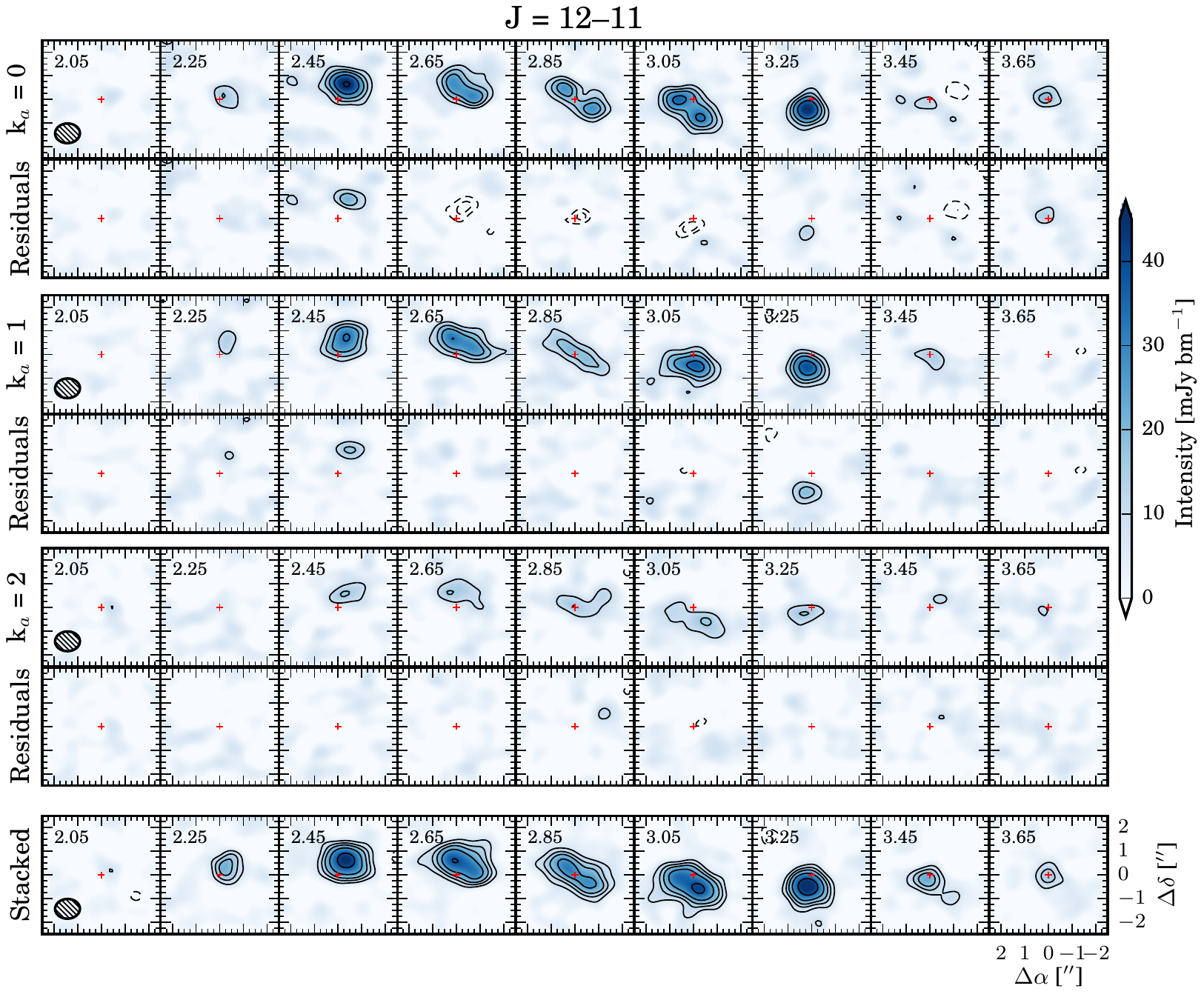}
    \caption{{\small Channel maps of the observed CH$_3$CN J=12--11 transitions and residuals after subtracting synthesized observations from the full network chemical model. The observations were imaged with 0.2~km~s$^{-1}$ channel spacing and all panels share the same intensity scale. Contours are [3,5,7,...]$\times\sigma$, where $\sigma$=3.2~mJy~km~s$^{-1}$. The synthesized beam is shown in the left panel of each row.} 
    \label{Channel Maps 1}}
    \end{figure*}

\begin{figure*}[ht!]
    \centering
    \includegraphics[width=0.8\textwidth]{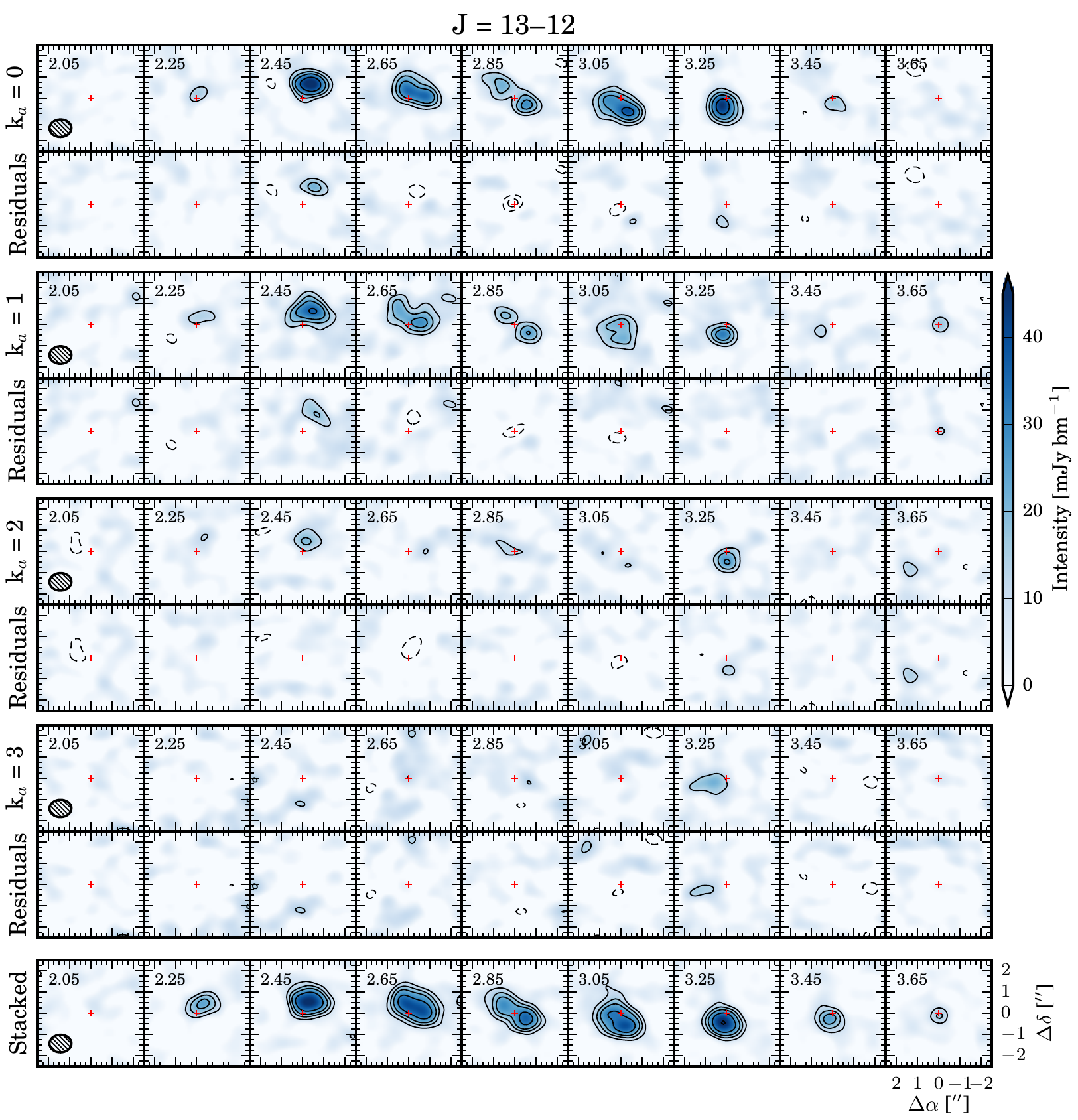}
    \caption{{\small Same as Fig. \ref{Channel Maps 1}, but for the J=13--12 k-ladder of CH$_3$CN.}
    \label{Channel Maps 2}}
    \end{figure*}

    Channel maps of the observed CH$_3$CN transitions are shown in Figs. \ref{Channel Maps 1} and \ref{Channel Maps 2}, generated from the image cubes described in \S2 at 0.2~km~s$^{-1}$. Residuals after subtracting the synthesized observations from the full network chemical model described in \S3.3 are shown in the figures as well, alternating rows with the observations. Although these residuals are small, a trend is seen where the model over-produces emission in the central channels and under-produces emission at $\pm$0.4~km~s$^{-1}$. This can likely be attributed to the small deviation from Keplerian rotation at small radii in TW Hya \citep{Rosenfeld_2012}, which we only partially able to account for by slightly increasing the overall inclination of our model (from 6$\degree$ to 8$\degree$).
    
    To investigate the distribution of CH$_3$CN at small radii, we additionally stacked the transitions within each k-ladder to improve the signal-to-noise ratio (SNR). The filter responses of each transition were used as an estimate of their inherent SNR, and applied as stacking weights when the measurement sets were combined in the uv-plane. The stacked measurement sets were identically imaged to the individual transitions, and show evidence for emission up to 0.8~km~s$^{-1}$ from the systemic velocity. This velocity corresponds to a radius of $\sim$16~AU, assuming a stellar mass of 0.8~M$_{\odot}$ and inclination of 7 degrees. The data are therefore compatible with the presence of a depression in CH$_3$CN surface density at small radii, but constrain the outer radius of such a potential feature to be less than 16~AU.

\clearpage
    
\section{MCMC Fit Covariance}
\begin{figure}[ht!]
    \centering
    \includegraphics[width=0.5\columnwidth]{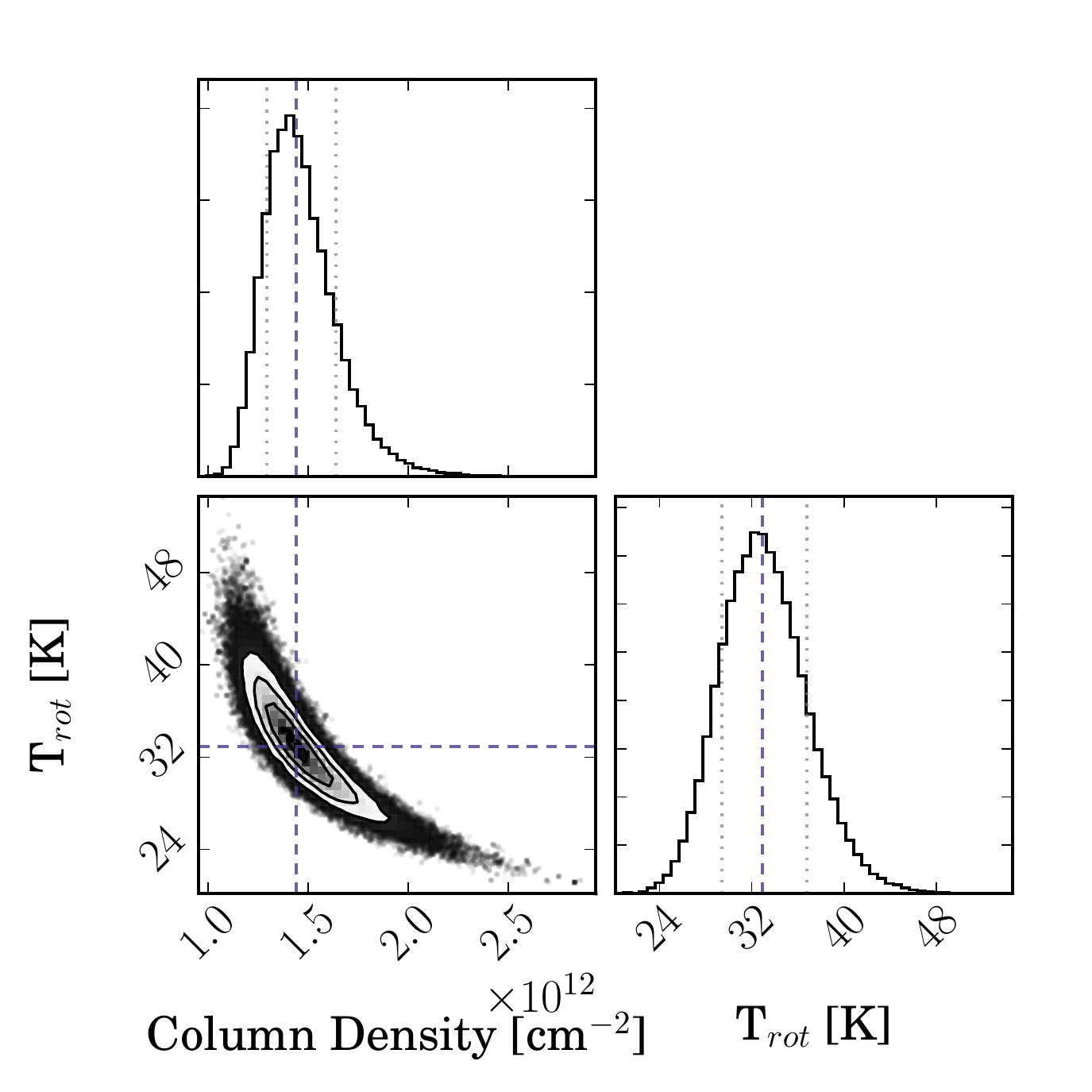}
    \caption{{\small A corner plot showing the posterior probability distributions and covariances for the rotational diagram fit shown in Fig. \ref{Figure 4}. The 16th, and 84th percentiles for each parameter are shown in dotted grey, with the 50th percentile shown in dashed blue.}
    \label{corner fig}}
    \end{figure}

    A corner plot showing the posterior probability distributions and covariances for the rotational diagram fit shown in Fig. \ref{Figure 4} is shown in Fig. \ref{corner fig}. Similar covariances are observed for the fit to each radial bin in Fig. \ref{Figure 5}.
    
\end{document}